\documentclass[10pt]{article}
\pdfoutput=1
\usepackage[usenames]{color} %used for font color
\usepackage{amssymb} %maths

\usepackage{amsmath} %maths
\usepackage{float}
\usepackage[pdftex]{graphicx}
\usepackage[utf8]{inputenc} %useful to type directly diacritic characters
\usepackage{geometry}
\usepackage[numbers,sort]{natbib}
\usepackage{url}
\usepackage{hyperref}
\usepackage[bottom]{footmisc}

\begin{document}

\pdfoutput=1

\title{Data-Driven Precision Luminosity Measurements with Z Bosons at the LHC and HL-LHC}
\author{}
\maketitle
\begin{center}
Jakob Salfeld-Nebgen, Daniel Marlow

Princeton University
\end{center}
A method to measure integrated luminosities at the LHC using Z bosons without theoretical cross section input is discussed. The main uncertainties and the prospects for precision luminosity measurements using this method are outlined.

\section{Introduction}

Luminosity measurements allow one to relate the production cross section of a physics process to
the number of observed events via
\begin{equation}
N_{\rm events} = \sigma L,
\end{equation}
where $N_{\rm events}$ is the observed number of events, $\sigma$ is the cross section, and $L$ is the integrated luminosity of the dataset under study.   Precise measurements of cross sections can be used to probe for new physics.   For example, parton interactions are subject to perturbative higher order corrections in the standard model and can possibly contain small corrections from new heavy states entering through loop diagrams.  A general relation between the precision of a coupling measurement and sensitivity to new physics is given by
\begin{equation}
\frac{\Delta g}{ g_{\rm SM}} \propto \frac{1}{\Lambda^{2}},
\end{equation}
which provides the expected deviation from the standard model coupling $g_{\rm SM}$, and the corresponding cross section, due to new interactions at a higher energy scale $\Lambda$.  

Uncertainties in the luminosity measurement at the LHC limit the absolute precision of cross section measurements and hence the ability to probe small differences in coupling strengths.   
It is therefore important to perform precise luminosity measurements and to explore different approaches to doing so.  The challenge will be particularly daunting at the High-Luminosity LHC (HL-LHC), where processes will be measured with excellent statistical precision, but under data-taking conditions where the mean number of  proton-proton collisions per bunch crossing (pileup or PU) is as large as 200. 
  
Here we discuss a method to use Z boson yields to perform a precise luminosity measurement, 
independent of theoretical assumptions.  In this method, the Z cross section is measured at low instantaneous luminosities soon after the completion of the van-der-Meer (vdM) beam-beam scans that are used for absolute luminosity scale calibration\cite{vanderMeer:296752}.
  The luminosity uncertainty will, in this period, mainly be the one on the normalization, typically about 1\% - 1.5\%\cite{Aaboud:2016hhf,CMS:2017sdi}.  The muon efficiency can be assumed to be measured with a precision similar to that of the nominal Z cross section measurements performed under nominal data-taking conditions (see e.g. Ref.~\cite{Aad:2016jkr}).   The luminosity in production running is then determined using the ratio of efficiency-corrected Z boson yields measured in the same fiducial volume as in the low instantaneous luminosity run.  With this procedure the luminosity uncertainty is composed of two components:

\begin{itemize}
\item the uncertainty on the absolute luminosity scale calibration
\item uncorrelated muon efficiency uncertainties between two muon efficiency measurements, one performed at moderate/high pileup conditions characteristic of data taking and the other at low pileup conditions used for the vdM scan.   The correlated part will cancel in the measurement procedure.
\end{itemize}

Luminosity measurements based on Z boson yields were, for example, discussed in Ref.~\cite{Dittmar:1997md} and are being used during Run-2 for monitoring of the relative LHC luminosity distribution to the ATLAS and CMS experiments\cite{lpcweb}.  

\section{Luminosity Measurements}
\label{LUMMeas}
The experiments at the LHC have employed the van-der-Meer scan method for absolute luminosity scale calibration.  Luminosity measurements are performed using a variety of algorithms, such as the counting of clusters in pixel detectors, collision vertices, particle tracks,  and the monitoring of subdetector channel occupancies.  The ATLAS and CMS experiments have measured integrated luminosities with a precision of 1.9\% - 2.5\%.  The LHCb collaboration achieved a luminosity measurement with 1.2\% uncertainty during LHC`s Run-1~\cite{Aaij:2014ida}, due to its excellent beam-gas imaging technique and low instantaneous luminosity data-taking conditions.  Table~\ref{uncLumRef} summarizes the normalization and integration uncertainties of the ATLAS collaboration at 8 TeV and the CMS collaboration at 13 TeV during the 2016 data-taking period, as an example.

\begin{table}[H]
\centering
\begin{tabular}{c || c | c }
 & Luminosity Scale & Calibration Transfer   \\
\hline
\hline
  ATLAS 8 TeV~\cite{Aaboud:2016hhf}  & 1.2\%  & 1.5\%     \\
  CMS 13 TeV 2016~\cite{CMS:2017sdi}  & 1.4\%   & 1.9\%    \\
   \end{tabular}
\caption{Example of luminosity uncertainties from the ATLAS and CMS collaborations subdivided into the calibration and transfer (integration).  The CMS measurement is preliminary. \label{uncLumRef}} 
\end{table}

Uncertainties in the luminosity measurement are typically subdivided into uncertainties on the absolute luminosity scale calibration derived from the vdM scan calibration procedure at low instantaneous luminosity, and the uncertainty due to possible changes of the calibration constants that result from data taking conditions at high instantaneous luminosities (calibration transfer).    
Uncertainties in the absolute luminosity scale are actively being improved.   The LHCb collaboration has measured the absolute luminosity scale with a precision of 1.1\% for their 8 TeV dataset.    It is thus expected that uncertainties in the calibration transfer to nominal high pileup data-taking conditions will dominate the luminosity uncertainty.   In particular, accurate calibration transfer requires that the methods used to measure the luminosity be highly linear, since the pileup values for production running can be one or two orders of magnitude higher than the pileup values employed during vdM scans.    Another important source of transfer calibration uncertainty is long-term calibration drift, which can occur, for example, as the result of detector aging. 

\section{Muon Efficiencies and Z Cross Section Measurements}
Z production cross sections can be measured with good precision at the LHC using the dimuon final state.  The ATLAS and CMS experiments have performed such measurements at 7, 8 and 13 TeV center of mass energies\cite{Chatrchyan:2014mua,Aaboud:2016zpd}.

Apart from the uncertainty on the luminosity measurement, the cross section measurements are affected by the uncertainty on the muon efficiencies.  Muon efficiencies are typically measured using the tag-and-probe method\cite{Khachatryan:2010xn} on a well known standard model dimuon decay resonance, for example the Z peak.

The ATLAS collaboration has performed a study of systematic uncertainties entering the muon efficiency measurement at 13 TeV~\cite{Aad:2016jkr}.  In this study, the uncertainty of the muon efficiency measurement is estimated to be less than 1\% for muons with transverse momentum larger than 20 GeV across the full pseudorapidity acceptance of the ATLAS detector. 
Uncertainties associated with the background estimation and the closure of the tag-and-probe method in simulation are the leading components.

For the proposal in this note, the correlation between uncertainties of muon efficiency measurements performed on independent datasets at the same center of mass energy but at different pileup conditions is important.   While detailed uncertainty correlation studies between muon efficiency measurements remain to be performed, one can expect that the uncertainty at the same center of mass energy and using the same muon selection and reconstruction methods is to a large extent correlated. 
 
In Ref.~\cite{Aaboud:2016zpd}, the uncertainties in efficiencies due to the trigger and isolation requirements are taken to be correlated for measurements performed at different center of mass energies and data-taking conditions. The uncertainties on the reconstruction and selection are taken to be uncorrelated for different center of mass energies, due to changes in the algorithms during Run-1 and Run-2. 
 
Efficiency effects due to high pileup data-taking conditions are taken from simulation, where the number of proton-proton collisions per bunch crossing is simulated according to the observed pileup spectrum.   
Muon scale factors in Ref.~\cite{Aaboud:2016zpd} are observed to be pileup independent, leading to the conclusion that pileup effects are well modeled in the simulation. In Ref.~\cite{Aaboud:2016zpd}, the uncertainty in the $Z\rightarrow \mu\mu$ cross section measurement at 13 TeV due to pileup mismodelling is evaluated to be $<$ 0.1\% and the total uncertainty component due to muon efficiencies is evaluated to be 0.8\%.

\section{Luminosity Measurement using Z Bosons}
As discussed in Section~\ref{LUMMeas}, the uncertainty in the luminosity measurement can be subdivided into the uncertainty entering the absolute luminosity scale calibration and the uncertainty associated with the calibration transfer or integration.   At low instantaneous luminosities, where occupancies are low and the detector response is linear, the luminosity can be measured with good precision.   Under these data-taking conditions the uncertainty on the luminosity measurement can be assumed to be similar to that of the vdM scan calibration procedure, as highlighted in Table~\ref{uncLumRef}.

In 2017, the LHC performed a low PU data-taking period of one week, where around 200~pb$^{-1}$ of data was recorded.  With a fiducial cross section of about 700~pb for $Z\rightarrow \mu\mu$ production processes at 13 TeV, the statistical uncertainty on Z boson yields can be expected to be 0.3\%.

The total integrated luminosity $L_{\rm tot}$ for a given dataset can now be measured by performing the ratio of efficiency corrected Z boson yields recorded in this low pileup dataset $nZ_{\rm low-PU}$ and the full dataset $nZ_{\rm tot}$

\begin{equation}
L_{\rm tot} = L_{\rm low-PU} \frac{nZ_{\rm tot}}{nZ_{\rm low-PU}},
\end{equation} 
where $L_{\rm low-PU}$ is the luminosity in the low pileup dataset. The relative uncertainty is now computed to be
\begin{equation}
\label{masterUncEq}
\delta^2 L_{\rm tot} = \delta^2 L_{\rm low-PU} + \delta^2 \epsilon^{Z\rightarrow\mu\mu}_{\rm tot} + \delta^2 \epsilon^{Z\rightarrow\mu\mu}_{\rm low-PU} - 2\rho\delta\epsilon^{Z\rightarrow\mu\mu}_{\rm tot}\delta\epsilon^{Z\rightarrow\mu\mu}_{\rm low-PU},
\end{equation} 
where $\rho$ denotes the correlation of uncertainties of muon efficiencies $\epsilon_{\rm tot}$, $\epsilon_{\rm low-PU}$ in the two datasets and $\delta$ denotes the relative uncertainty.  As discussed, 
$\delta L_{\rm low-PU}$ is about 1\% - 1.5\%. If the correlation between the uncertainties on the muon efficiencies is 100\%, these uncertainties will cancel. An upper and lower bound on the expected uncertainty can be derived by combining the uncertainty in the $Z\rightarrow\mu\mu$ cross section measurement due to muon efficiencies of 0.8\%~\cite{Aaboud:2016zpd} and the uncertainties in the luminosity scale of 1.2\%, as shown in Table~\ref{uncLumRef}. Based on these recent results, the expected uncertainty in the luminosity  is thus 1.2\% - 1.8\%, where in the upper bound the uncertainty in the muon efficiencies is assumed to be uncorrelated an thus added in quadrature twice to the 1.4\% luminosity scale uncertainty shown in Table~\ref{uncLumRef}.

While one can assume that the systematic effects of muon efficiencies so far studied are the same at low and high pileup, there may be pileup-dependent effects associated with the tag-and-probe method.  In this case Eq.~\ref{masterUncEq} can be rewritten as:
\begin{equation}
\label{masterUncEq2}
\delta^2 L_{\rm tot} = \delta^2 L_{\rm low-PU} + \delta^2 \epsilon^{Z\rightarrow\mu\mu}_{\rm PU},
\end{equation} 
where $\delta \epsilon^{Z\rightarrow\mu\mu}_{\rm PU}$ denotes the relative uncertainty on the $Z\rightarrow\mu\mu$ reconstruction efficiency due to pileup. 
There are at least two contributions to this uncertainty.  One is the simulation of pileup in samples that typically rely on proper modeling of soft non-perturbative QCD effects in proton-proton collisions.  This component can, for example, be studied in a data-driven procedure by embedding simulated hard-scattering events into a data sample recorded by minimum or zero bias triggers.  The second is the modeling of the detector used to reconstruct simulated events in the collider experiments. 

Two particular cases are mentioned below, where $\delta L_{\mathrm{transfer}}$ denotes the luminosity uncertainty associated with the calibration transfer, shown in the right-hand column of Table~\ref{uncLumRef}.
\begin{itemize}
\item $\delta \epsilon^{Z\rightarrow\mu\mu}_{\rm PU}<\delta L_{\mathrm{transfer}}$: the luminosity measurement based on the method discussed in this document is more precise than the conventional luminosity measurements.
\item $\delta \epsilon^{Z\rightarrow\mu\mu}_{\rm PU}>\delta L_{\mathrm{transfer}}$: the uncertainty on the luminosity is not the leading systematic uncertainty, but rather the luminosity measurement can be used to decrease the uncertainty on $Z\rightarrow\mu\mu$ reconstruction efficiencies.
\end{itemize}
The method discussed here and conventional luminosity measurements can also be combined to improve the overall precision on the luminosity estimate for other cross section measurements.

\section{Conclusions}
A method to measure integrated luminosity values using Z boson yields is discussed.  In contrast to previous approaches, no theoretical assumptions enter the measurements.  The method exploits the excellent precision of luminosity measurements for low pileup data-taking periods and the correlation between uncertainties of muon efficiency measurements. 

Implementation of the method will require further detailed studies of muon efficiency uncertainties and relies on excellent understanding of lepton efficiency measurements using the tag-and-probe method.  Similarly, the method can be used for any other physical production mechanism with sufficient rate and well studied reconstruction efficiencies.

\section*{Acknowledgements}
The authors thank Aram Apyan for useful comments on a previous draft of this document.

\bibliographystyle{unsrt} 
\bibliography{biblum4}

\begin{thebibliography}{10}

\bibitem{vanderMeer:296752}
S.~van~der Meer.
\newblock {Calibration of the effective beam height in the ISR}.
\newblock Technical Report CERN-ISR-PO-68-31, CERN, Geneva, 1968.

\bibitem{Aaboud:2016hhf}
{ATLAS Collaboration}.
\newblock {Luminosity determination in pp collisions at $\sqrt{s}$ = 8 TeV
  using the ATLAS detector at the LHC}.
\newblock {\em Eur. Phys. J. $\bf{C 76}$ (2016) 653}.

\bibitem{CMS:2017sdi}
CMS Collaboration.
\newblock {CMS Luminosity Measurements for the 2016 Data Taking Period}.
\newblock (CMS-PAS-LUM-17-001), 2017.

\bibitem{Aad:2016jkr}
ATLAS Collaboration.
\newblock {Muon reconstruction performance of the ATLAS detector in
  proton–proton collision data at $\sqrt{s}$ =13 TeV}.
\newblock {\em Eur. Phys. J.}, C76(5):292, 2016.

\bibitem{Dittmar:1997md}
M.~Dittmar, F.~Pauss, and D.~Zurcher.
\newblock {Towards a precise parton luminosity determination at the CERN LHC}.
\newblock {\em Phys. Rev.}, D56:7284--7290, 1997.

\bibitem{lpcweb}
{LPC} performance plots.
\newblock \url{https://lpc.web.cern.ch/cgi-bin/plots.py}.
\newblock Accessed: 2018-05-28.

\bibitem{Aaij:2014ida}
{LHCb \vspace{0mm}Collaboration, R. Aaij et al}.
\newblock {Precision luminosity measurements at LHCb}.
\newblock {\em 2014 JINST $\bf{9}$ P12005}.

\bibitem{Chatrchyan:2014mua}
CMS Collaboration.
\newblock {Measurement of inclusive W and Z boson production cross sections in
  pp collisions at $\sqrt{s}$ = 8 TeV}.
\newblock {\em Phys. Rev. Lett.}, 112:191802, 2014.

\bibitem{Aaboud:2016zpd}
ATLAS Collaboration.
\newblock {Measurements of top-quark pair to $Z$-boson cross-section ratios at
  $\sqrt s = 13, 8, 7$ TeV with the ATLAS detector}.
\newblock {\em JHEP}, 02:117, 2017.

\bibitem{Khachatryan:2010xn}
CMS Collaboration.
\newblock {Measurements of Inclusive $W$ and $Z$ Cross Sections in $pp$
  Collisions at $\sqrt{s}=7$ TeV}.
\newblock {\em JHEP}, 01:080, 2011.

\end{thebibliography}

\end{document}